\documentclass{article}

\usepackage{graphicx}
\usepackage{listings}
\usepackage{lstbayes}
\usepackage{xcolor}
\usepackage{xr}
\usepackage{glossaries}
\usepackage{imakeidx}
\usepackage{hyperref}
\usepackage{cleveref}

\usepackage{pdflscape}
\usepackage{afterpage}
\usepackage{authblk}

\makeindex

\newcommand{\ewbankslope}{2.1}
\newcommand{\frenchslope}{2.09}
\newcommand{\uiaaslope}{2.13}
\newcommand{\verminslope}{3.17}

\newglossaryentry{onsight}{name=onsight, description={is a successful ascent of a route on the very first attempt without weighting the rope, and without having any information about how the route should be climbed. If information is known about the route from other climbers, or by observing other climbers on the route, then it becomes a flash}}

\newglossaryentry{flash}{name=flash, description={is a successful ascent of a route or boulder problem on the very first attempt. If on a lead route then success means the climber did not weight the rope. It differs from onsight in that any amount of information about the route, including beta and observation of other climbers on the route is available to the climber when attempting a flash}}

\newglossaryentry{redpoint}{name=redpoint, description={is a successful ascent of a route or boulder problem on any attempt subsequent to the first attempt. Successful means that the rope was not weighted and no other aid than the natural rock features was used. If a successful ascent is logged as a redpoint it implies that climber has made previous attempts and/or ascents}}

\newglossaryentry{hangdog}{name=hangdog, description={is an ascent of a route in which the climber attained the top of the climb, but did not do so in a style that was deemed successful, because they weighted the rope. When working to redpoint a hard route, a climber will often climb it hangdog as an intermediate stage towards success}}

\newglossaryentry{send}{name=send, description={is a successful ascent of a route or boulder problem}}

\makeglossaries

\begin{document}

\title{Bayesian inference of the climbing grade scale}
\author[1,2]{Alexei Drummond}
\author[2]{Alex Popinga}
\affil[1]{School of Computer Science, University of Auckland, Auckland, New Zealand}
\affil[2]{School of Biological Sciences, University of Auckland, Auckland, New Zealand}

\date{}
\maketitle

\abstract{Climbing grades are used to classify a climbing route based on its perceived difficulty, and have come to play a central role in the sport of rock climbing. Recently, the first statistically rigorous method for estimating climbing grades from whole-history ascent data was described \cite{scarff2020estimation}, based on the dynamic Bradley-Terry model for games between players of time-varying ability. In this paper, we implement inference under the whole-history rating model using Markov chain Monte Carlo and apply the method to a curated data set made up of climbers who climb regularly. We use these data to get an estimate of the model's fundamental scale parameter $m$, which defines the proportional increase in difficulty associated with an increment of grade. We show that the data conform to assumptions that the climbing grade scale is a logarithmic scale of difficulty, like decibels or stellar magnitude.

We estimate that an increment in Ewbank, French and UIAA climbing grade systems corresponds to $\ewbankslope{}$, $\frenchslope{}$ and $\uiaaslope$ times increase in difficulty respectively, assuming a logistic model of probability of success as a function of grade. Whereas we find that the Vermin scale for bouldering (V-grade scale) corresponds to a $\verminslope$ increase in difficulty per grade increment. In addition, we highlight potential connections between the logarithmic properties of climbing grade scales and the psychophysical laws of Weber and Fechner.  

\section*{Author summary}

In this paper, we ask the question: ``What does an increment of the climbing grade really mean?''  Climbing grades originated as a way of classifying the difficulty of ascending particular routes based on subjective determination by the climbers.  However, the grades exhibit strongly quantitative behaviour, which we analyse by forming a model that explains the grade of a route as a function of the number of sessions that a climber of a certain ability will take to successfully climb the route. We propose assigning grades to climbers as well as to routes and then expressing the expected number of failures before success as a logarithmic function of the difference between the climber's grade and the grade of the route.  We find that by using this approach, we can estimate both the grade of the climber and the slope of the climbing grade scale itself from a logbook of attempts that includes all of the climbers' successes and failures to ascend a set of routes.  Using data from a popular public online climbing log book (\url{thecrag.com}), we find that for three different climbing grade scales that an increment of a sport climbing grade corresponds to slightly more than a doubling of the expected number of failures before successfully climbing the route.

\section*{Introduction}

Climbing grades play a central role in the sport of rock climbing \cite{delignieres1993psychophysical,draper2015comparative}. There are a number of commonly employed grading systems used by sport climbers globally to classify climbing routes into grades that increase with difficulty.  In North America, the three-part Yosemite decimal system (YDS) is employed.  In many parts of Europe, the French sport grading system is used.  In Australia and NZ, climbers use the simple numerical Ewbank grading scale. The Ewbank scale is open-ended and begins at 1. Most amateur climbers will be able to climb routes up to about grade 18, while the most difficult routes currently graded under the Ewbank system are grade 35. 

For lead climbing at the advanced end of the spectrum (i. e., from about Ewbank 23 onwards) there is an almost one-to-one correspondence between these three major grading systems \cite{draper2015comparative}, (see Table \ref{table1}).  However, this simple relationship is not universally accepted, and a fourth major grade system (UIAA) is usually assumed to have wider intervals in most conversion tables \cite{draper2015comparative}.

For bouldering, the two common grading systems are the Vermin scale and the Fontainbleau scale. The consensus view as that these two scales are also have a correspondence in the higher grades, so that V11 $\equiv$ 8A, V12 $\equiv$ 8A+, $\dots$, V17 $\equiv$ 9A.

\begin{table}
\centering
\begin{tabular}{| c | c | c |}
  \hline			
  {\bf Ewbank} & {\bf French sport} & {\bf YDS} \\
  \hline			
  23 & 7a & 5.11d \\
  24 & 7a+ & 5.12a \\
  25 & 7b & 5.12b \\
  26 & 7b+ & 5.12c \\
  27 & 7c & 5.12d \\
  28 & 7c+ & 5.13a \\
  29 & 8a & 5.13b \\
  30 & 8a+ & 5.13c \\
  31 & 8b & 5.13d \\
  32 & 8b+ & 5.14a \\
  33 & 8c & 5.14b \\
  34 & 8c+ & 5.14c \\
  35 & 9a & 5.14d \\
  36 & 9a+ & 5.15a \\
  37 & 9b & 5.15b \\
  38 & 9b+ & 5.15c \\
  39 & 9c & 5.15d \\
  \hline  
\end{tabular}
\caption{The correspondence between three of the major sport climbing grade systems in the range of grades relevant to advanced and elite climbers. For a more complete table see \cite{draper2015comparative}. }
\label{table1}
\end{table}

Climbing grades were developed subjectively by different climbing communities around the world to classify climbing routes by difficulty. Yet, for difficult routes, these grading systems appear to be converging. They are also quite predictive of the chance of a climber of known ability to successfully climb a new route.  This has led to the idea that there may be a quantitative law that underlies climbing grades \cite{delignieres1993psychophysical}, perhaps similar to the psychophysical Weber-Fechner law \cite{1834pulsu,fechner1860}. The Weber-Fechner law states that the human perception of sensation intensity is proportional to the logarithm of the stimulus.  In climbing, the grade represents the perceived difficulty (usually as determined by the first ascensionist). In controlled conditions it has been shown that climbing grades are proportional to the logarithm of some objective measures of difficulty \cite{delignieres1993psychophysical}. 

Many physical and physiological determinants of climbing performance have been investigated \cite{balavs2012hand, balavs2014relationship, mackenzie2020physical}. If it exists, objective climbing difficulty must be a very complex quantity, but in certain situations it can perhaps be approximated by some relatively simple measurable proxy.  One example of a measurable proxy exists in electromyographic stimulus of the {\it flexor digitorum profundus} (FDP) \cite{delignieres1993psychophysical}.  A climbing route that is primarily difficult because of the ``crimpy'' nature of its holds could be well characterised by such a proxy, since the FDP is the primary muscle involved in closing the fingers into the ``crimp'' grip position used by climbers to hold small ledges.  Another example of a measureable proxy for assessing the complex difficulty of a climb was investigated in a recent study that found that shoulder strength was the most significant factor when considering performance on indoor lead-climbing routes \cite{mackenzie2020physical}, which are often set to be more ``athletic'' and less ``fingery'' (mostly relating to the crimps) than outdoor routes.

Recently, a new statistical approach to objectively estimate climbing route difficulty using whole-history ascent data of the climbing community was described \cite{scarff2020estimation}. This approach adapts an earlier method for Bayesian inference of player ratings in two-player games \cite{coulom2008whole}, by recasting the sport of climbing as a game between the climber and the route. Such models have a long history dating back to at least Zermelo \cite{zermelo1929berechnung}, whose model was rediscovered decades later by Bradley and Terry \cite{bradley1952rank}. These models have been developed, extended, and generalised to estimate the time-varying skill of players of board games like Chess \cite{Elo1978, glickman1999rating} and Go \cite{coulom2008whole}, computer games like Starcraft \cite{maystre2019pairwise}, and sports such as tennis \cite{maystre2019pairwise} and basketball \cite{maystre2019pairwise}.

By applying this two-player game framework to sport climbing, climbers can be graded in the same way that routes are graded. An attempt to climb a route is a game between the climber and the route. If the climber and the route have the same grade, then the game will be fair, and there will be even odds that the outcome of the game is success on the part of the climber. This model introduces a scale parameter $m$ to describe how the probability of success changes as a function of the difference between the grade of the climber and that of the route.

Below, we investigate the suitability of climber ascent data to be described by the Bradley-Terry model.  We then apply Bayesian MCMC inference to estimate the time-varying skill of a set of sport climbers along with the fundamental parameter $m$, which, by establishing a probability of success on the part of the climber, can also be used to describe the increase in difficulty associated with an increment in the climbing grade.  We compare our estimates with earlier work.  We also examine the assumptions of the whole-history method, leading us to suggest some improvements that take into consideration common self-reporting behaviours that exist in popular public log books of climbing ascents that can significantly affect the accuracy of the model.

\section*{Model}

\subsection*{Ascent Data}

Let $\mathcal{D} = \{(c_i, r_i, t_i, y_i) : i \in [N]\}$ be a dataset of $N$ outcomes of sport climbing ascents.  Each ascent $i$, occurs on date $t_i$ and involves climber $c_i \in \{1,\dots, N_c\}$ attempting an ascent of route $r_i \in \{1,\dots, N_r\}$ with outcome $y_i \in \{0,1\}$. An outcome of $y=1$ represents a successful `clean' ascent of the climb, while $y=0$ represents a failure to make a clean ascent (including ``\gls{hangdog}'', failed attempt, or retreat). 

\subsection*{The Bradley-Terry model}

Following previous work  \cite{scarff2020estimation} we apply the dynamic Bradley-Terry model \cite{zermelo1929berechnung,bradley1952rank} to ascent data. We define the grade, $C(t)$, of a climber at time $t$ to be the grade for which they would have an even chance of climbing a route cleanly of that grade on any given attempt. A 7a climber, thus defined, would have a 50\% chance of successfully ``flashing'' (succeeding to climb it on the first attempt with information on the route available) a 7a route. 

Following previous work \cite{scarff2020estimation} we will posit the following equation for the probability of a successful attempt:

\begin{equation}
\Pr(y = 1) = p_{send} = \frac{e^{mC(t)}}{e^{mC(t)} + e^{mR}}
\end{equation}

where $C(t)$ is the grade of the climber and $R$ is the grade of the route, and $m$ is a slope parameter that defines the proportional increase in difficulty associated with an increment of the grade of the route. This equation can be re-arranged to reveal that it represents a logistic function of $p$ versus the grade difference $C(t)-R$:

\begin{equation}
p_{send} = \frac{1}{e^{m(R-C(t))} + 1} = \textrm{logit}^{-1}(mC(t)-mR),
\label{logisticp}
\end{equation}

where $\textrm{logit}^{-1}$ is the cumulative distribution of the logistic distribution, also known as the inverse logistic function:

\begin{equation}
\textrm{logit}^{-1}(x) = \frac{1}{1 + e^{-x}}
\end{equation}

Therefore estimating a climber's grade based on the outcomes of their ascent attempts is equivalent to a logistic regression with one independent variable. The success or failure of the attempt is the outcome, and the difference in the grade of the climber and the route is the independent variable. Assuming a logistic function is the same as stating that the log-odds of success is a linear function of the difference in the grades.



Assuming the route grade is known, a climber's grade, $C(t)$, can be seen as a scaled version of the intercept of the logistic regression. So defined, it does not have to be a whole number. The Ewbank grading scale is attractive for considering fractional climbing grades as it is purely numerical. So we could consider a climber to have grade 29.4 on the Ewbank scale, meaning that they have a better than even chance of flashing a grade 29 route and a worse than even chance of flashing a grade 30 route.

The grade of a climber is not static. The ability of a climber tends to increase rapidly in the first few months and years in the sport. Besides that, form waxes and wanes with injury, seasons, training intensity, et cetera. So any mechanism for estimating a climber's grade over a period of time must take account of the fact that the climber's ``effective'' grade will, in general, increase or decrease over time.

Following previous authors  \cite{coulom2008whole,scarff2020estimation} we consider a climber's grade to change through time according to a Wiener process, so that the prior distribution on $C(t+1)$ is:

\begin{equation}
C(t+1) \sim \mathcal{N}(C(t), w^2)  
\end{equation}

If we know the grade of a climber, we can predict their expected performance on a route of some other grade, including grades and routes they haven't tried. 


If a climber has a probability $p_{send}$ of success on any particular attempt of a route, then it follows that the expected number of failed attempts before the redpoint (first ``clean'' success) will equal to the odds-ratio of failure:

\begin{equation}
 E\left[a\right]=\frac{1-p_{send}}{p_{send}}
 \end{equation} 
 
 So if the climber is climbing a route for which $p_{send}=0.5$ (i. e., a fair game) then they would expect to climb that route in an average of 2 attempts. They would flash it with a 50\% probability, take two goes with a 25\% probability (i. e., 1 fail, 1 success), 3 goes with a 12.5\% probability, et cetera.

Realising this relationship between $p_{send}$ and $ E\left[a\right]$ means that we can convert between one and the other. So if a climber fails $9$ times for every send when climbing routes of a given grade, then we can compute the probability  $p_{send} = \frac{1}{E(a)+1} = 0.1$.

\subsection*{Session grades}

In the above model, we have considered a game to be defined by a single attempt by a climber to climb a route cleanly, (otherwise known as a ``send" or ``to send the route", wherein a climber moves from the ground to the top using only the rock or plastic holds assigned to the route, in the case of indoor climbing; the gear and rope are only there for safety and do not assist in the climbing on a ``send"). An attempt is also known within the climbing community as a ``tie-in'' or a ``go''. A tie-in is the act of a climber tying the rope to their harness, then attempting to send the route, and either succeeding or returning to the ground and untying in order to rest before another go.  During a tie-in, a climber may choose to continue to ascend the rock even after a failure to send has occurred (by falling or intentionally resting on the rope or gear); they may wish to practice moves or positions on the rock, place more gear for the next attempt, etc.

A single climber may have multiple tie-ins for a given route in a single day or ``session" (a period of time, usually lasting from a couple of hours up to a whole day, during which a climber is trying to succeed in sending a route).  Unfortunately, few climbers log the result of every tie-in during a session, and in these cases it is more realistic to define the climbing game by the outcome of each session's attempts.

If the game is defined on a session-basis, then the outcome of the game is the best result that the climber achieved during the session. The climber wins if they eventually achieved a clean ascent during the session, no matter how many attempts they made to achieve the clean ascent. Although this form of ascent data is more granular (at most one ascent outcome per route per day, assuming a day is equivalent to a session), it may be preferred because a larger fraction of the climbing community logs data that conforms to this definition. 

We would anticipate that a climber's session grade, thus defined, is greater than their flash grade. 

%
%
%
%
%
%

\section*{Methods}

We analysed 20 datasets across three countries (New Zealand, Australia, Germany) using data from The Crag (\url{http://thecrag.com}), a popular online climbing logbook. In doing so we evaluate four different grading systems (Ewbanks, French sport, UIAA, V-scale) and three styles of climbing involving different types of gear or no gear (Sport, Trad, Bouldering). Table \ref{datatable} summarises the datasets.

\afterpage{%
    \clearpage
    \thispagestyle{empty}
    \begin{landscape}
\begin{table}[ht]
\centering
\fontsize{9pt}{10pt}\selectfont
\begin{tabular}{llrrrrrrrllr}
  \hline
{\bf country} & {\bf gear (style)} & {\bf climbers} & {\bf ascents} & {\bf slope} & {\bf hpd.lower} & {\bf hpd.upper} & {\bf min.ascents} & {\bf min.failures} & {\bf grade.type} & {\bf game} & {\bf time} \\ 
  \hline
Australia & Boulder & 100 & 25805 & 2.96 & 2.85 & 3.07 &  30 &   1 & V-grade & attempt & 6468 \\ 
  Australia & Boulder & 100 & 25344 & 3.03 & 2.91 & 3.15 &  30 &   1 & V-grade & session & 5752 \\ 
  Australia & Sport & 100 & 52947 & 2.24 & 2.21 & 2.28 &  30 &   1 & Ewbanks & attempt & 4142 \\ 
  Australia & Sport & 100 & 48679 & 2.33 & 2.29 & 2.37 &  30 &   1 & Ewbanks & session & 4094 \\ 
  Australia & Sport+Trad & 100 & 61931 & 2.13 & 2.10 & 2.16 &  30 &   1 & Ewbanks & attempt & 4905 \\ 
  Australia & Sport+Trad & 100 & 57512 & 2.18 & 2.15 & 2.21 &  30 &   1 & Ewbanks & session & 4729 \\ 
  Australia & Trad & 100 & 14929 & 2.02 & 1.96 & 2.09 &  30 &   1 & Ewbanks & attempt & 4899 \\ 
  Australia & Trad & 100 & 14479 & 2.01 & 1.95 & 2.08 &  30 &   1 & Ewbanks & session & 4324 \\ 
  New Zealand & Boulder &  15 & 1049 & 3.35 & 2.71 & 4.27 &  30 &   1 & V-grade & attempt &  41 \\ 
  New Zealand & Boulder &  15 & 1027 & 3.35 & 2.68 & 4.32 &  30 &   1 & V-grade & session &  40 \\ 
  New Zealand & Sport &  89 & 10027 & 2.19 & 2.11 & 2.27 &  30 &   1 & Ewbanks & attempt & 1561 \\ 
  New Zealand & Sport &  89 & 9679 & 2.22 & 2.13 & 2.31 &  30 &   1 & Ewbanks & session & 2397 \\ 
  New Zealand & Sport+Trad &  98 & 11389 & 2.15 & 2.08 & 2.22 &  30 &   1 & Ewbanks & attempt & 2597 \\ 
  New Zealand & Sport+Trad &  98 & 11024 & 2.17 & 2.09 & 2.26 &  30 &   1 & Ewbanks & session & 2344 \\ 
  New Zealand & Trad &   8 & 524 & 1.80 & 1.57 & 2.11 &  30 &   1 & Ewbanks & attempt &  47 \\ 
  New Zealand & Trad &   8 & 517 & 1.77 & 1.55 & 2.07 &  30 &   1 & Ewbanks & session &  30 \\ 
  Germany & Sport &   8 & 521 & 2.08 & 1.79 & 2.46 &  30 &   1 & French & attempt &  42 \\ 
  Germany & Sport &   8 & 483 & 2.09 & 1.79 & 2.49 &  30 &   1 & French & session &  25 \\ 
  Germany & Sport & 100 & 19497 & 2.12 & 2.06 & 2.18 &  30 &   1 & UIAA & attempt & 5026 \\ 
  Germany & Sport & 100 & 18435 & 2.14 & 2.08 & 2.21 &  30 &   1 & UIAA & session & 4949 \\ 
   \hline
\end{tabular}
\caption{Summary of Bayesian analyses performed} 
\label{datatable}
\end{table}

    \end{landscape}
    \clearpage
}

We describe two analyses in more detail. The first highlighted analysis uses self-reported data on whole-history (meaning climber recorded every attempt) of ascents for 100 Australian climbers of differing abilities to estimate the fundamental slope parameter for the Ewbank climbing grade scale for sport climbing. We repeated this analysis using the same criteria for selection (see Supplementary Material) for climbers in New Zealand, resulting in an additional dataset of 89 climbers.

\begin{figure}
\centering
\includegraphics[width=\textwidth]{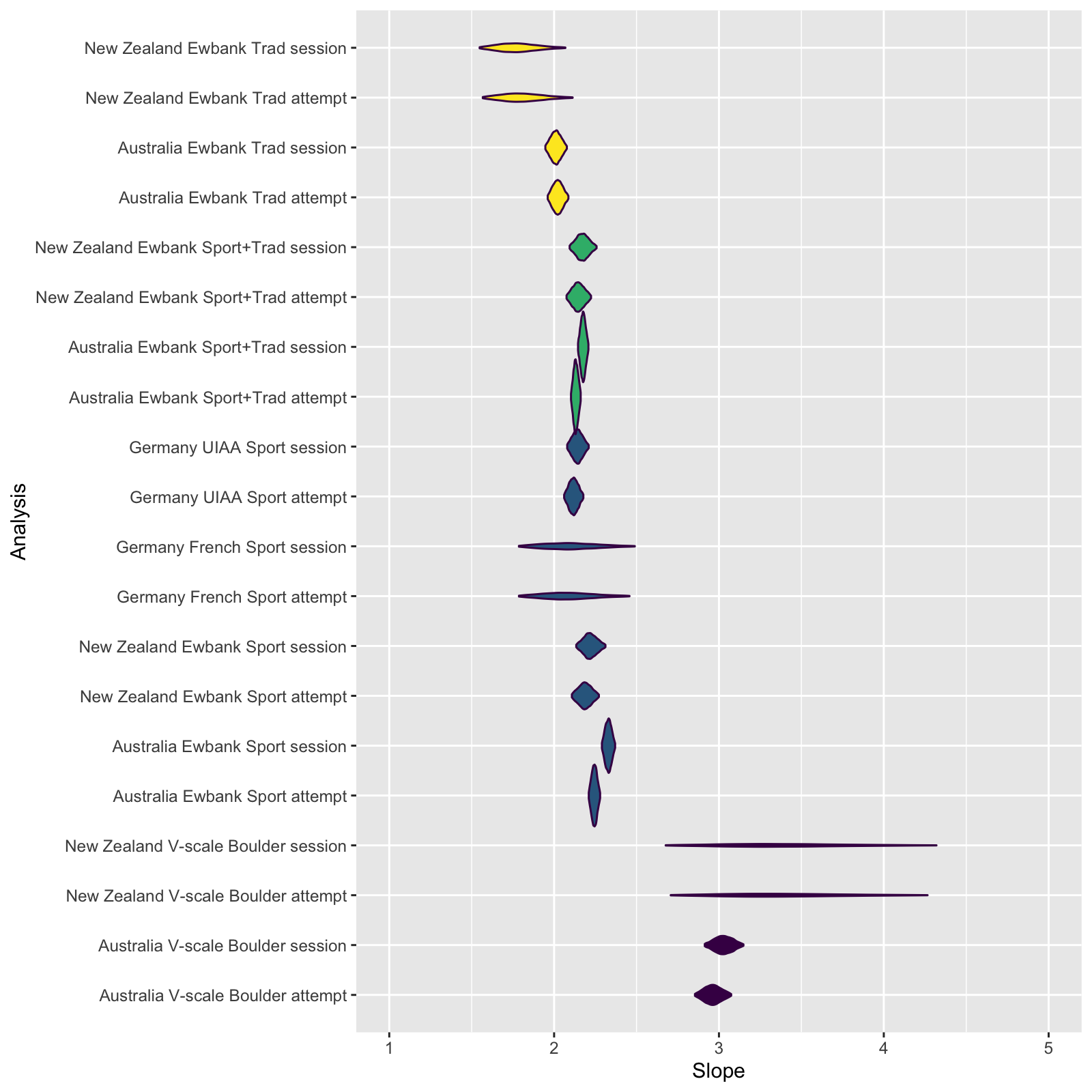}
\caption{\small The posterior estimate of the slope parameter ($d=e^m$) for each of the 20 analyses.}
\label{finalplot}
\end{figure}

The estimated slopes of all 20 analyses are presented in Figure \ref{finalplot}.

In the second set of analyses, we use whole-community successful ascent data from New Zealand and Australia. 

\subsection*{Estimating the slope of difficulty for the Ewbank climbing grade scale}

For the Australia-Sport analysis we used self-reported whole-history of ascents for $n=100$ climbers of differing abilities. All of these log books were self-reported on \url{http://thecrag.com}, a popular online climbing logbook. The total number of ascents analysed was $N=48,679$. Most climbers do not self-report every attempt. To investigate the bias caused by climbers that don't log every failure we also converted all logbooks to a session logging form, in which only the single most successful ascent was chosen for each climber on each route on each day that climbing occurred. In addition we only considered climbers that had at least 30 attempts, with at least 1 explicit failure (i. e., at least one \gls{hangdog}, attempt, retreat or working). The analysis took 1hr and 8 minutes on an iMac 3.6 GHz 8-Core Intel Core i9.

We implemented full Bayesian MCMC inference of the dynamic Bradley-Terry model in Stan \cite{carpenter2017stan}. This code is available to the public domain at \url{https://github.com/alexeid/climbing-grades}.
The Stan model employed is shown in \cref{supp-listing1} (Supplementary Information).

We used this model to co-estimate the fundamental slope parameter $m$ and the grade of each climber in monthly windows during 60 months between 1st August 2016 and 1st August 2021. For the purpose of this analysis we assumed that the assigned grades for the routes attempted were correct.

\subsection*{Estimating the slope of total successful ascents versus grade in community-wide ascent data}

We followed O'Neill \cite{oneill2002} in analysing total successful ascent data, as a complementary approach to understanding grade difficulty. We downloaded the total number of successful ascents by grade for New Zealand and Australia from \url{http://thecrag.com}, for both bouldering and sport climbing on 20th July 2021. For sport climbing we chose to include ascents with tick types: \gls{redpoint}, \gls{flash} and \gls{onsight}. For bouldering we chose to include tick types \gls{send}, \gls{flash} and \gls{onsight}. The purpose was to chose tick types that strongly implied a clean send on the part of the ascensionist (i. e., represented accomplishing success of the full difficulty of the boulder problem or route).



\section*{Results}

\subsection*{Estimating the slope of difficulty for the Ewbank climbing grade scale}

Figure \ref{aus_ascents} shows the estimated grade through time plot between August 2016 and July 2021 inclusive for 100 climbers that fulfilled our selection criteria for the Australia Sport data set. The second panel reports the posterior distribution of the slope parameter which was jointly estimated.
The estimate of the $m$ parameter was 0.85 [0.83, 0.86], which is equivalent to 2.33 [2.29, 2.37] times more failed attempts per success per grade increment. 

The mean estimate of the $m$ parameter from a simple log-linear regression of each climber individually assuming no change in ability of the analysed period was 0.65 (see Figure \ref{supp-fig1}; Supplementary Information). 

The median number of explicit fails per climber (i. e., \gls{hangdog}, attempt, working, retreat), was 126 (range: 18-642) or as a fraction 29.9\% (interquartile range 4.7\%- 72.0\%) of ascents.
\begin{figure}
\centering
\includegraphics[width=\textwidth]{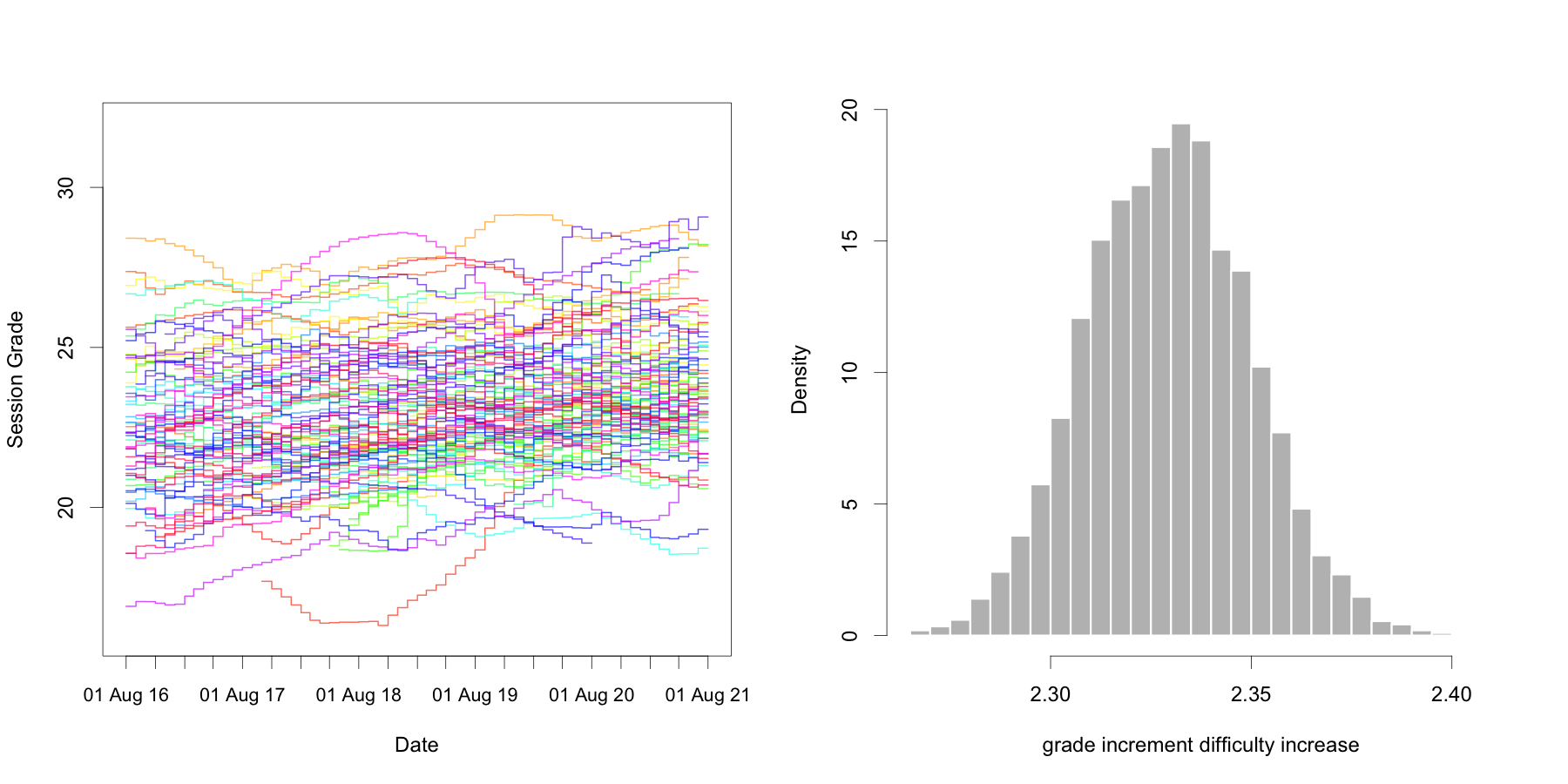}
\caption{\small The posterior estimate of each Australian climber's grade ($n=100$) through time and the posterior distribution of the proportional increase in difficulty per grade increment $d = e^m$.}
\label{aus_ascents}
\end{figure}

Figure \ref{nz_ascents} shows the estimated grade through time plot between August 2016 and July 2021 inclusive for 89 climbers that fulfilled our selection criteria for the New Zealand Sport data set. The second panel reports the posterior distribution of the slope parameter which was jointly estimated.

The estimate of the $m$ parameter was 0.8 [0.76, 0.84], which is equivalent to 2.22 [2.13, 2.31] times more failed attempts per success per grade increment. 

The mean estimate of the $m$ parameter from a simple log-linear regression of each climber individually assuming no change in ability of the analysed period was 0.52 (see Figure \ref{supp-fig2}; Supplementary Information).

\begin{figure}
\centering
\includegraphics[width=\textwidth]{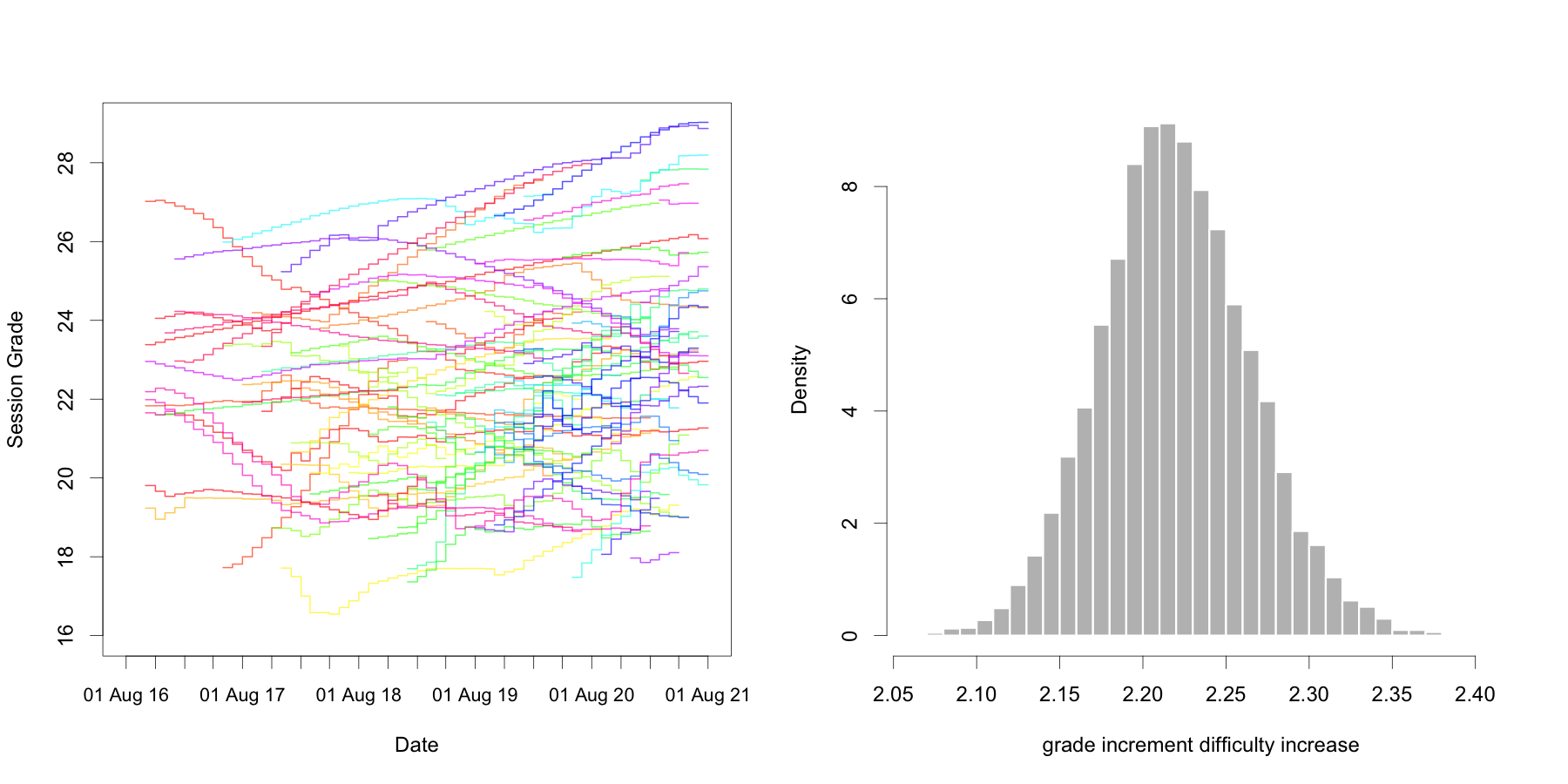}
\caption{\small The posterior estimate of each New Zealand climber's grade through time and the posterior distribution of the proportional increase in difficulty per grade increment $d = e^m$.}
\label{nz_ascents}
\end{figure}

\subsection*{Estimating the slope of total successful ascents versus grade in community-wide ascent data}

Figure \ref{supp-fig1} shows the whole-community counts of successful ascents by grade for Australian and New Zealand for both sport climbs and bouldering. The slope of these curves range between 0.7 and 0.79, conforming remarkably well with the more detailed Bayesian analysis above. This data is not affected by bias in under-reporting of failed attempts, but will be affected by systematic variation in the number of attempts made by the community as a whole at each grade.

\begin{figure}
\centering
\includegraphics[width=\textwidth]{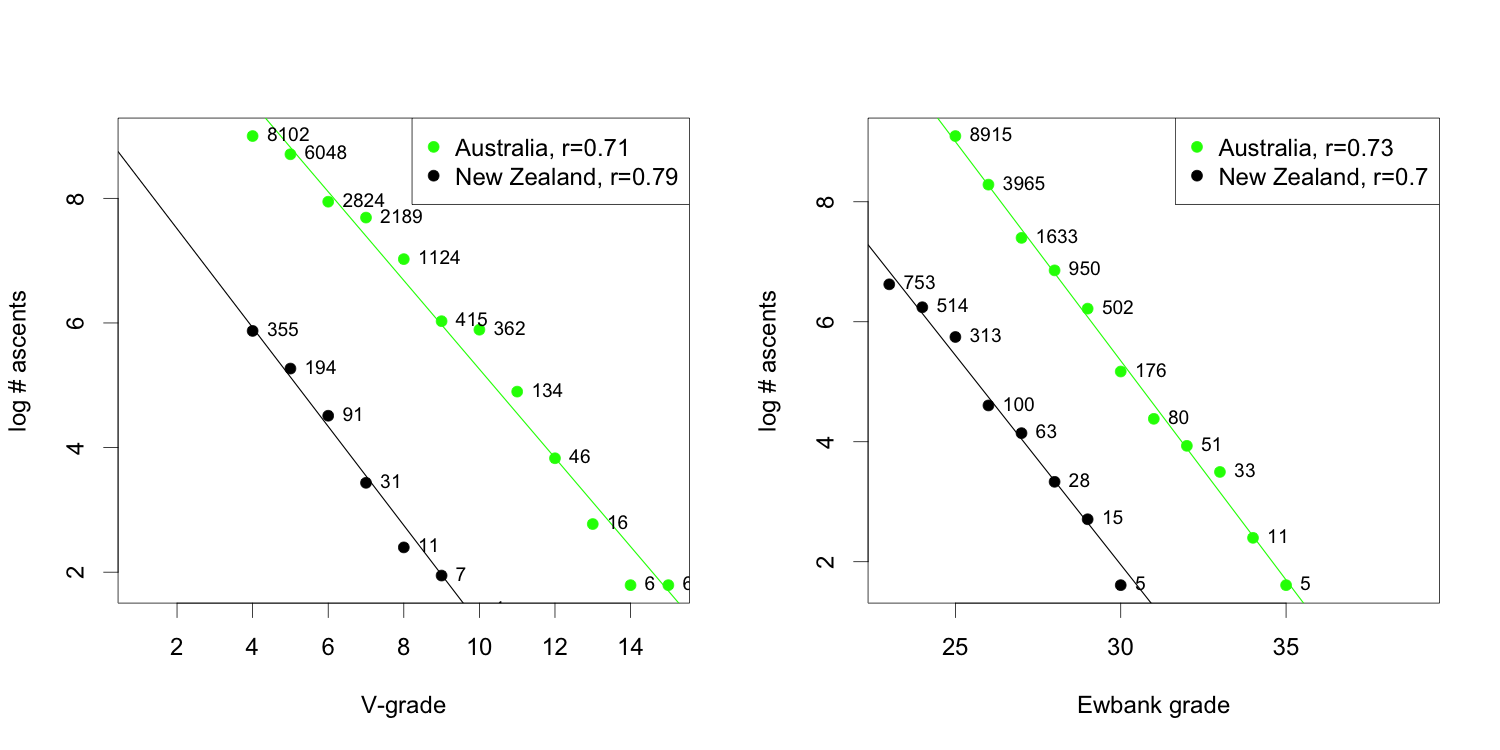}
\caption{\small The relationship between grade and the log of the number of successful sends for two styles of climbing (bouldering and sport climbing) in two countries, based on whole climbing community statistics on \url{http://thecrag.com}}
\label{fig.oneill}
\end{figure}

\section*{Discussion and Conclusion}

We have endeavoured to bring together various pieces of community knowledge about the mathematics of climbing grades and reframe them in the light of recent work on grade estimation using whole-history ascent data.

Our main contribution is to focus on quantifying the magnitude of the increase in difficulty representing by an increment in climbing grade. We provide evidence that for the Ewbank, French sport and UIAA grading systems this increment is slightly more than a doubling in difficulty for each grade, as measured by average number of failed sessions required before achieving success. For the V-grade system we find that the increment in difficulty is slightly more than a tripling in difficulty for each grade.

This is quite a surprising quantitative law to arise from what many have assumed to be a subjective process of grading climbs. Whereas climbers do argue whether a climb is too hard or too ``soft'' for the grade, implying its true grade is off by one, it is much more rare for disagreements about the grade of a route to span more than two consecutive grades. Typically there is one grade representing the consensus, and a neighbouring grade the alternative. This makes sense if the grade scale is in fact a logarithmic scale of difficulty.  

An open question is how these subjective grading systems settled in to such a rigid quantitative law. One hypothesis worth further investigation is that a logarithmic scale is optimal for magnitude estimation and therefore arises naturally in many settings \cite{portugal2011weber}. In many different domains it appears that the perception of a stimulus varied logarithmically with its absolute magnitude. If so, why did these (sport climbing) grading systems settle on a slope so close to 2? Is there an information theoretic reason for this? Furthermore, why do bouldering grades seem to have a definitely wider scale? Is there a natural reason for this difference related to the smaller number of moves on boulder problems, or the greater number of attempts possible in a session?

One of the key data preparation approaches used here that differs from previous whole-history efforts is to use per-session data. This is a form of aggregation in which only the most successful attempt for each climber on each route on each day is retained and all other attempted ascents are removed from the data set. This caters for a common form of self-reporting that reports the best effort during each day of climbing. We refer to the grade estimated for a climber in this way as the {\it session grade}. Since it is possible to try a route 2 or more times in a single session, a climber's {\it session grade} should be higher than the {\it flash grade}. 

The grade of a climb defines its overall difficulty for the average climber in appropriate conditions. However there are many factors besides the grade that can affect the probability of success. Many routes are best climbed in certain conditions, including time of day, season and weather. Direct sun on a route is normally not conducive to optimal performance and many climbers prefer shade, cool rock, a slight breeze and low humidity for the best chances of success. Besides conditions, routes come in many styles; a climber may specialise in slabs (positive angle), vertical or overhanging routes. Depending on the rock type and crag, different hold types may predominate (pockets, slopers, crimps, underclings, jugs), and the climbing may be more ``cruxy''/``bouldery'' with rests, or of a ``pumpy''/``resistance'' nature more suited to climbers with endurance. Finally, some routes, or their cruxes, are easier to climb for certain body types or sizes. A crux move may be easier for a tall climber because of their extra reach, or easier for a shorter climber because they can fit their body into a small space between holds (called ``fitting in the box''). Routes for which the difficulty varies greatly depending on the size of the climber are sometimes termed ``morpho''. Suffice to say, there are many factors that could be taken into account in order to elaborate the simple model described here. Arguably the most obvious would be to include the style of the route (e.g. slab, vertical, steep/overhung, or cruxy versus pumpy) and estimate a correction factor for each climber on each style. 

The most questionable detail of the model presented is the idea that the probability of a successful ascent remains the same after each previous attempt of the route, i.e., practising a particle route does not improve the climber's chance of success. This is clearly a poor assumption, so at best the probability of success implied by the Bradley-Terry model should be considered as some sort of ``effective'' probability, averaged over different levels of practice. The true underlying probability of success per attempt is probably increasing with practice for most routes. The problems with developing a model that admits learning are at least two-fold: (i) we expect some routes are more amenable to practice than others, and (ii) it is unclear what functional form the expected improvement in probability per attempt should take.

Besides details of the model, there are also potential problems related to how climbers choose the routes to climb (selection bias), and also which ascent attempts they decide to log (under-reporting failures). 

Selection bias is the propensity of climbers to chose routes that are at the easier end of the grade, or at least routes that suit their climbing style (rather than choosing a random climb at the grade). There are two reasons why we think that selection bias shouldn't have a large impact on the estimate of the climbing grade slope parameter. Firstly, climbers are probably less selective at lower to middle grades in their range, and the slope is derived from the full range of grades that a climber attempts. Secondly, if a climber is selective at all grades, such a bias doesn't change the slope, but instead changes the intercept, because in each grade the result will be to select from the lower end of the grade, which will still result in the same slope. The main result of such selection bias will be an overestimate the grade of the climber by up to 1 grade. If the climbers only select a certain style of climb, then their grade estimate is only relevant to that style of climb. So selection bias will overestimate a climber's ability a bit (either overall, or for styles they don't try), but it won't have a big effect on the fundamental grade scale parameter estimate. The worst case scenario will be if all climbers only climb relatively hard routes in grades low in their range and relatively easy routes from grades high in their range. This would mean the true range in the x-axis of the regression for each climber is actually up to a grade less than what is used to compute the slope. So the estimated slope will be flatter than the true slope. This extreme scenario would flatten the slope by approximately $(k-1)/k$, where $k$ is the difference between the highest grade and the lowest grade. In our data set the interquartile range for $k$ is 8-10. So the worst case scenario if everybody is maximally biased in this way is an underestimate of the slope of about 10\%.  

Selective logging of failures is a more significant concern. It is clear that most climbers don't log all failures. It seems likely that failures on easy routes would be more embarrassing than failures on harder routes. So if this motivation is commonly in action, then harder routes will appear even harder than they actually are, since the failures are underreported for easier routes. This would lead to an overestimation of the slope of the grade scale. We see some evidence that this might be the case when comparing our results to data of very experienced climbers known to log every attempt (data not shown). For climbers known to log every attempt, the slope tends to be slight less than 2, rather than slightly greater than 2 (data not shown).

Suffice to say, there is still much work to do. Foremost in our mind is the need to adapt recently developed whole-history inference methods to account for biases in public repository data in a much more rigorous way than pursued here. This will require a better understanding of the differing ways that climbers approach self-reporting of climbing ascents. It seems likely that climber that logs their own ascent and attempts are likely to be susceptible to various biases and to follow differing conventions, depending on their purpose for making a public log book of their ascents. Data-driven approaches to learning about these differing conventions and biases in order to classifying climbers by their logging approach is an obvious next step.

\section*{Acknowledgement}

The authors would like to thank Simon Dale and Ulf Fuchslueger from \url{thecrag.com}, for providing access to the theCrag API (\url{https://www.thecrag.com/en/article/api}) so that the public ascent data could be downloaded programmatically for the analyses produced in this work. In addition we thank Simon Dale, Dr Joseph Heled, Daniel Krippner, Dr Michael Matschiner, John Palmer and Dr Tim Vaughan for helpful discussions on earlier versions of this manuscript.

\printglossaries

\bibliography{climbing}

\begin{thebibliography}{10}
\providecommand{\url}[1]{#1}
\csname url@samestyle\endcsname
\providecommand{\newblock}{\relax}
\providecommand{\bibinfo}[2]{#2}
\providecommand{\BIBentrySTDinterwordspacing}{\spaceskip=0pt\relax}
\providecommand{\BIBentryALTinterwordstretchfactor}{4}
\providecommand{\BIBentryALTinterwordspacing}{\spaceskip=\fontdimen2\font plus
\BIBentryALTinterwordstretchfactor\fontdimen3\font minus
  \fontdimen4\font\relax}
\providecommand{\BIBforeignlanguage}[2]{{%
\expandafter\ifx\csname l@#1\endcsname\relax
\typeout{** WARNING: IEEEtran.bst: No hyphenation pattern has been}%
\typeout{** loaded for the language `#1'. Using the pattern for}%
\typeout{** the default language instead.}%
\else
\language=\csname l@#1\endcsname
\fi
#2}}
\providecommand{\BIBdecl}{\relax}
\BIBdecl

\bibitem{scarff2020estimation}
D.~Scarff, ``Estimation of climbing route difficulty using whole-history
  rating,'' \emph{arXiv preprint arXiv:2001.05388}, 2020.

\bibitem{delignieres1993psychophysical}
D.~Deligni{\`e}res, J.-P. Famose, C.~Th{\'e}paut-Mathieu, P.~Fleurance
  \emph{et~al.}, ``A psychophysical study of difficulty rating in rock
  climbing,'' \emph{International Journal of Sport Psychology}, vol.~24, pp.
  404--404, 1993.

\bibitem{draper2015comparative}
N.~Draper, D.~Giles, V.~Sch{\"o}ffl, F.~Konstantin~Fuss, P.~Watts, P.~Wolf,
  J.~Bal{\'a}{\v{s}}, V.~Espana-Romero, G.~Blunt~Gonzalez, S.~Fryer
  \emph{et~al.}, ``Comparative grading scales, statistical analyses, climber
  descriptors and ability grouping: International rock climbing research
  association position statement,'' \emph{Sports Technology}, vol.~8, no. 3-4,
  pp. 88--94, 2015.

\bibitem{1834pulsu}
\BIBentryALTinterwordspacing
\emph{De pulsu, resorptione, auditu et tactu: Annotationes anatomicae et
  physiologicae}, ser. De pulsu, resorptione, auditu et tactu: Annotationes
  anatomicae et physiologicae, 1834. [Online]. Available:
  \url{https://books.google.co.nz/books?id=j7CspFMOQTYC}
\BIBentrySTDinterwordspacing

\bibitem{fechner1860}
G.~T. Fechner, \emph{Elemente der psychophysik}.\hskip 1em plus 0.5em minus
  0.4em\relax Leipzig: Breitkopf und H{\"a}rtel, 1860.

\bibitem{balavs2012hand}
J.~Bal{\'a}{\v{s}}, O.~Pecha, A.~J. Martin, and D.~Cochrane, ``Hand--arm
  strength and endurance as predictors of climbing performance,''
  \emph{European Journal of Sport Science}, vol.~12, no.~1, pp. 16--25, 2012.

\bibitem{balavs2014relationship}
J.~Bal{\'a}{\v{s}}, M.~Pan{\'a}{\v{c}}kov{\'a}, B.~Strejcov{\'a}, A.~J. Martin,
  D.~J. Cochrane, M.~Kal{\'a}b, J.~Kodej{\v{s}}ka, and N.~Draper, ``The
  relationship between climbing ability and physiological responses to rock
  climbing,'' \emph{The Scientific World Journal}, vol. 2014, 2014.

\bibitem{mackenzie2020physical}
R.~MacKenzie, L.~Monaghan, R.~A. Masson, A.~K. Werner, T.~S. Caprez,
  L.~Johnston, and O.~J. Kemi, ``Physical and physiological determinants of
  rock climbing,'' \emph{International journal of sports physiology and
  performance}, vol.~15, no.~2, pp. 168--179, 2020.

\bibitem{coulom2008whole}
R.~Coulom, ``Whole-history rating: A bayesian rating system for players of
  time-varying strength,'' in \emph{International Conference on Computers and
  Games}.\hskip 1em plus 0.5em minus 0.4em\relax Springer, 2008, pp. 113--124.

\bibitem{zermelo1929berechnung}
E.~Zermelo, ``Die berechnung der turnier-ergebnisse als ein maximumproblem der
  wahrscheinlichkeitsrechnung,'' \emph{Mathematische Zeitschrift}, vol.~29,
  no.~1, pp. 436--460, 1929.

\bibitem{bradley1952rank}
R.~A. Bradley and M.~E. Terry, ``Rank analysis of incomplete block designs: I.
  the method of paired comparisons,'' \emph{Biometrika}, vol.~39, no. 3/4, pp.
  324--345, 1952.

\bibitem{Elo1978}
A.~E. Elo, \emph{The rating of chessplayers, past and present}.\hskip 1em plus
  0.5em minus 0.4em\relax New York: Arco Pub., 1978.

\bibitem{glickman1999rating}
M.~E. Glickman and A.~C. Jones, ``Rating the chess rating system,''
  \emph{CHANCE-BERLIN THEN NEW YORK-}, vol.~12, pp. 21--28, 1999.

\bibitem{maystre2019pairwise}
L.~Maystre, V.~Kristof, and M.~Grossglauser, ``Pairwise comparisons with
  flexible time-dynamics,'' in \emph{Proceedings of the 25th ACM SIGKDD
  International Conference on Knowledge Discovery \& Data Mining}, 2019, pp.
  1236--1246.

\bibitem{carpenter2017stan}
B.~Carpenter, A.~Gelman, M.~D. Hoffman, D.~Lee, B.~Goodrich, M.~Betancourt,
  M.~Brubaker, J.~Guo, P.~Li, and A.~Riddell, ``Stan: A probabilistic
  programming language,'' \emph{Journal of statistical software}, vol.~76,
  no.~1, pp. 1--32, 2017.

\bibitem{oneill2002}
\BIBentryALTinterwordspacing
T.~O'Neill. (2002, June) Grade theory? [Online]. Available:
  \url{https://web.archive.org/web/20020609071230/http://www.australianbouldering.com:80/table.html}
\BIBentrySTDinterwordspacing

\bibitem{portugal2011weber}
R.~Portugal and B.~F. Svaiter, ``Weber-fechner law and the optimality of the
  logarithmic scale,'' \emph{Minds and Machines}, vol.~21, no.~1, pp. 73--81,
  2011.

\end{thebibliography}


\begin{thebibliography}{1}
\providecommand{\url}[1]{#1}
\csname url@samestyle\endcsname
\providecommand{\newblock}{\relax}
\providecommand{\bibinfo}[2]{#2}
\providecommand{\BIBentrySTDinterwordspacing}{\spaceskip=0pt\relax}
\providecommand{\BIBentryALTinterwordstretchfactor}{4}
\providecommand{\BIBentryALTinterwordspacing}{\spaceskip=\fontdimen2\font plus
\BIBentryALTinterwordstretchfactor\fontdimen3\font minus
  \fontdimen4\font\relax}
\providecommand{\BIBforeignlanguage}[2]{{%
\expandafter\ifx\csname l@#1\endcsname\relax
\typeout{** WARNING: IEEEtran.bst: No hyphenation pattern has been}%
\typeout{** loaded for the language `#1'. Using the pattern for}%
\typeout{** the default language instead.}%
\else
\language=\csname l@#1\endcsname
\fi
#2}}
\providecommand{\BIBdecl}{\relax}
\BIBdecl

\bibitem{oneill2002}
\BIBentryALTinterwordspacing
T.~O'Neill. (2002, June) Grade theory? [Online]. Available:
  \url{https://web.archive.org/web/20020609071230/http://www.australianbouldering.com:80/table.html}
\BIBentrySTDinterwordspacing

\end{thebibliography}

\end{document}


\title{Supplementary Information for ``Bayesian inference of the climbing grade scale''}
\author{Alexei J Drummond and Alex Popinga}
\date{\today}
\maketitle

\section{Details of Bayesian analyses}

The Bayesian analysis was conducted in Stan. The full Stan code is provided in the listing below.

\subsection{Data}

The data was obtained from \url{thecrag.com} using the API (\url{https://www.thecrag.com/en/article/api}). Data was accessed on the 24th August 2021, at which time all public ascents in Australia were requested resulting in 33 JSON files (50,000 ascents per file) totalling 543,125,776 bytes.

A summary of the data processing is shown in Table \ref{table-data-processing-aus}.

\begin{table}[ht]
\centering
\begingroup\fontsize{9pt}{10pt}\selectfont
\begin{tabular}{r r p{11cm}}
  \hline
{\bf rows.in} & {\bf rows.out} & {\bf filter} \\ 
  \hline
1627548 & 1465494 & Exclude ascents with no date or no grade information. \\ 
  1465494 & 1429967 & Exclude artificial ascents \\ 
  1429967 & 432571 & Exclude gear styles: NA, Trad, Boulder, Top rope, Alpine, DWS, Unknown, Traverse, Aid, Ice, Via ferrata \\ 
  432571 & 432499 & Exclude trad ascent types: greenpoint, greenpointonsight \\ 
  432499 & 432493 & Exclude boulder ascent types: send, dab, repeat \\ 
  432493 & 431405 & Exclude non-ascent types: hit, target, mark \\ 
  431405 & 431286 & Keep only ascents graded with grade type: AU \\ 
  431286 & 431276 & Remove grades with value '--' \\ 
  431276 & 263360 & Remove ascents before 2016-08-01 \\ 
  263360 & 259229 & Remove ascents on or after 2021-08-01 \\ 
  259229 & 249527 & Remove ascents with no month information. \\ 
  249527 & 213608 & Remove ascents with grade less than 16 \\ 
  213608 & 170234 & Exclude ambiguous ascent types: tick, lead, leadsolo, second, toprope, aidsolo, ropedsolo \\ 
  170234 & 140216 & Keep climbers with at least 30 ascents, and at least 1 failed ascents. \\ 
  140216 & 52947 & Keep only the top 100 climbers ranked by total ascents\\ 
  52947 & 48679 & Keep only the best ascent on each route-climber-day \\ 
   \hline
\end{tabular}
\endgroup
\caption{Summary of data processing for analysis of Australia ascent data.} 
\label{table-data-processing-aus}
\end{table}

A summary of the data processing for the New Zealand data analysis is shown in Table \ref{table-data-processing-nz}

\begin{table}[ht]
\centering
\begingroup\fontsize{9pt}{10pt}\selectfont
\begin{tabular}{r r p{11cm}}
  \hline
{\bf rows.in} & {\bf rows.out} & {\bf filter} \\ 
  \hline
41546 & 38147 & Exclude ascents with no date or no grade information. \\ 
  38147 & 37928 & Exclude artificial ascents \\ 
  37928 & 26473 & Exclude gear styles: Ice, Boulder, Trad, Top rope, Alpine, NA, Unknown, DWS \\ 
  26473 & 26423 & Exclude trad ascent types: greenpoint, greenpointonsight \\ 
  26423 & 26423 & Exclude boulder ascent types: send, dab, repeat \\ 
  26423 & 26399 & Exclude non-ascent types: hit, target, mark \\ 
  26399 & 26383 & Keep only ascents graded with grade type: AU \\ 
  26383 & 26383 & Remove grades with value '--' \\ 
  26383 & 22022 & Remove ascents before 2016-08-01 \\ 
  22022 & 21908 & Remove ascents on or after 2021-08-01 \\ 
  21908 & 21198 & Remove ascents with no month information. \\ 
  21198 & 18583 & Remove ascents with grade less than 16 \\ 
  18583 & 13566 & Exclude ambiguous ascent types: tick, lead, leadsolo, second, toprope, aidsolo, ropedsolo \\ 
  13566 & 10027 & Keep climbers with at least 30 ascents, and at least 1 failed ascents. \\ 
  10027 & 9679 & Keep only the best ascent on each route-climber-day \\ 
   \hline
\end{tabular}
\endgroup
\caption{Summary of data processing for analysis of New Zealand ascent data.} 
\label{table-data-processing-nz}
\end{table}

\subsection{Model}

\begin{lstlisting}[language=Stan,caption={Stan model},label=listing1]
data {                          
  int<lower=1> C;            // number of climbers
  int<lower=1> N;            // number of ascents
  int<lower=1> P;            // number of pages
  int<lower=1> minPage[C];   // min page for each climber
  int<lower=1> maxPage[C];   // max page for each climber
  int<lower=0,upper=1> y[N]; // ascent success/failure
  int<lower=1> page[N];      // the time block (page) of each ascent
  int<lower=1> c[N];         // the climber of each ascent
  vector[N] x;               // route grade of each ascent
}
parameters {
  // mid-point intercept
  real climberGrade[C, P]; 
  
  // slope of increase in difficulty per grade increment
  real<lower=0.0> m;       
}
model {
  m ~ normal(0.69,0.3); // prior on slope
  for(j in 1:C) {
    // prior on grades up to and including first month with data
    for(i in 1:minPage[j]) {
      climberGrade[j, i]  ~ normal(18,5);    
    }
    // prior on grade in months after first that have data
    for(i in (minPage[j]+1):maxPage[j]) {
      climberGrade[j, i] ~ normal(climberGrade[j, i-1], 0.5);     
    }
    // prior on grades on months after data
    for(i in (maxPage[j]+1):P) {
      climberGrade[j, i]  ~ normal(18,5);    
    }
  }
  // likelihood
  for (i in 1:N) {
    y[i] ~ bernoulli_logit(m*(climberGrade[c[i], page[i]]-x[i])); 
  }    
}
\end{lstlisting}

\section{Assessing the suitability of the logistic model for climbing grade scale}

The average number of failed attempts per success is the odds-ratio of failure.  Under the logistic model there should be a linear relationship between the log-odds-ratio and the independent variable. Therefore, if the logistic model is a good fit, we would expect a linear relationship between the log of the number of failed attempts per success and the grade, for a climber of given ability.

\begin{figure}
\centering
\includegraphics[width=0.9\textwidth]{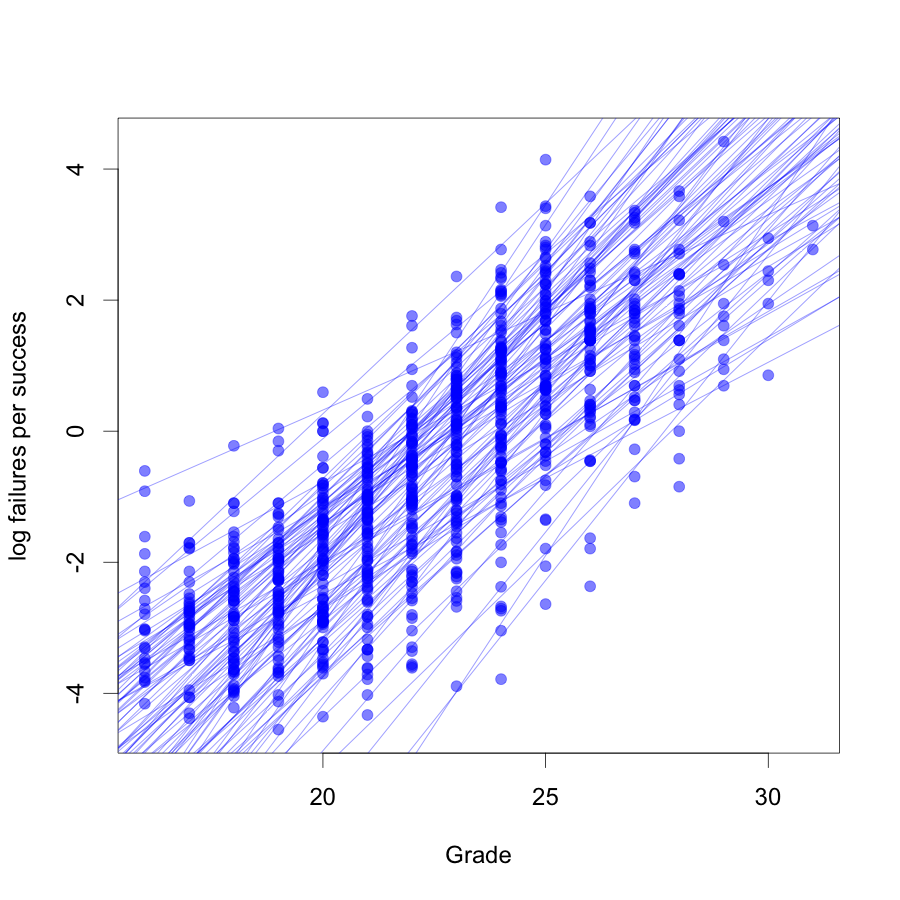}
\caption{\small The relationship between grade and the empirical log-odds-ratio of failure for the Australian climbers over a year of climbing. The point is plotted at log (\#failures / \#successes) for that grade. A regression is fit for each climber. Because there are a large number of climbers only the mean slope is reported in the main text. But it is clear that the slopes are rather consistent across climbers, despite the intercept varying by about 10 grades.}
\label{fig1}
\end{figure}

\begin{figure}
\centering
\includegraphics[width=0.9\textwidth]{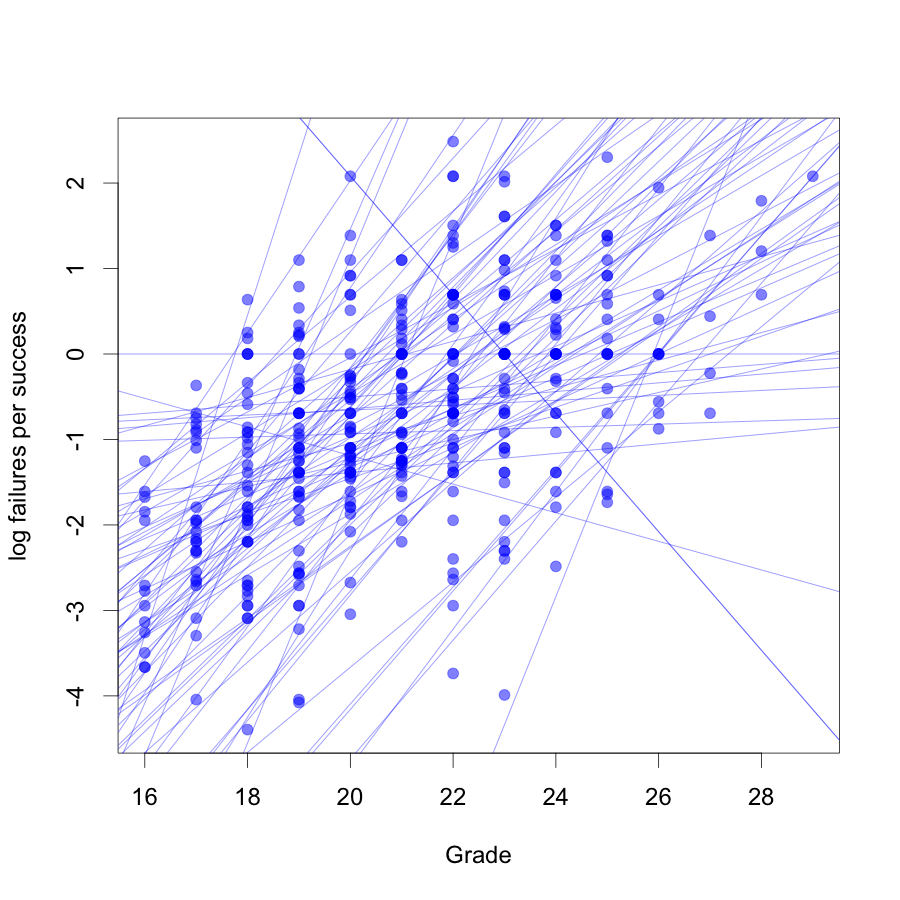}
\caption{\small The relationship between grade and the empirical log-odds-ratio of failure for the New Zealand climbers over a year of climbing. The number next to each point is the number of climbs of the grade successfully climbed over the course of the year. The point is plotted at log (\#failures / \#successes) for that grade. $m$ is the slope of the best fit, and $C_i$ is the estimated grade of the climber $i$ (i.e. the x value of the line corresponding to a log-odds-ratio of 0.}
\label{fig2}
\end{figure}

The relationship between grade and the number of failed attempts before redpoint for Australians is show in  Figure \ref{fig1}. The same analysis for New Zealand climbers is shown in Figure \ref{fig2}. 

Fitting a linear regression to the grade versus the logarithm of the number of failed attempts before redpoint allows us to get an approximate estimate for the $C$ and $m$ parameters, assuming $C(t)$ is constant over the period analysed.

\section{Estimates of flash grade and grade scale slope assuming whole-history data available}

\begin{figure}
\centering
\includegraphics[width=\textwidth]{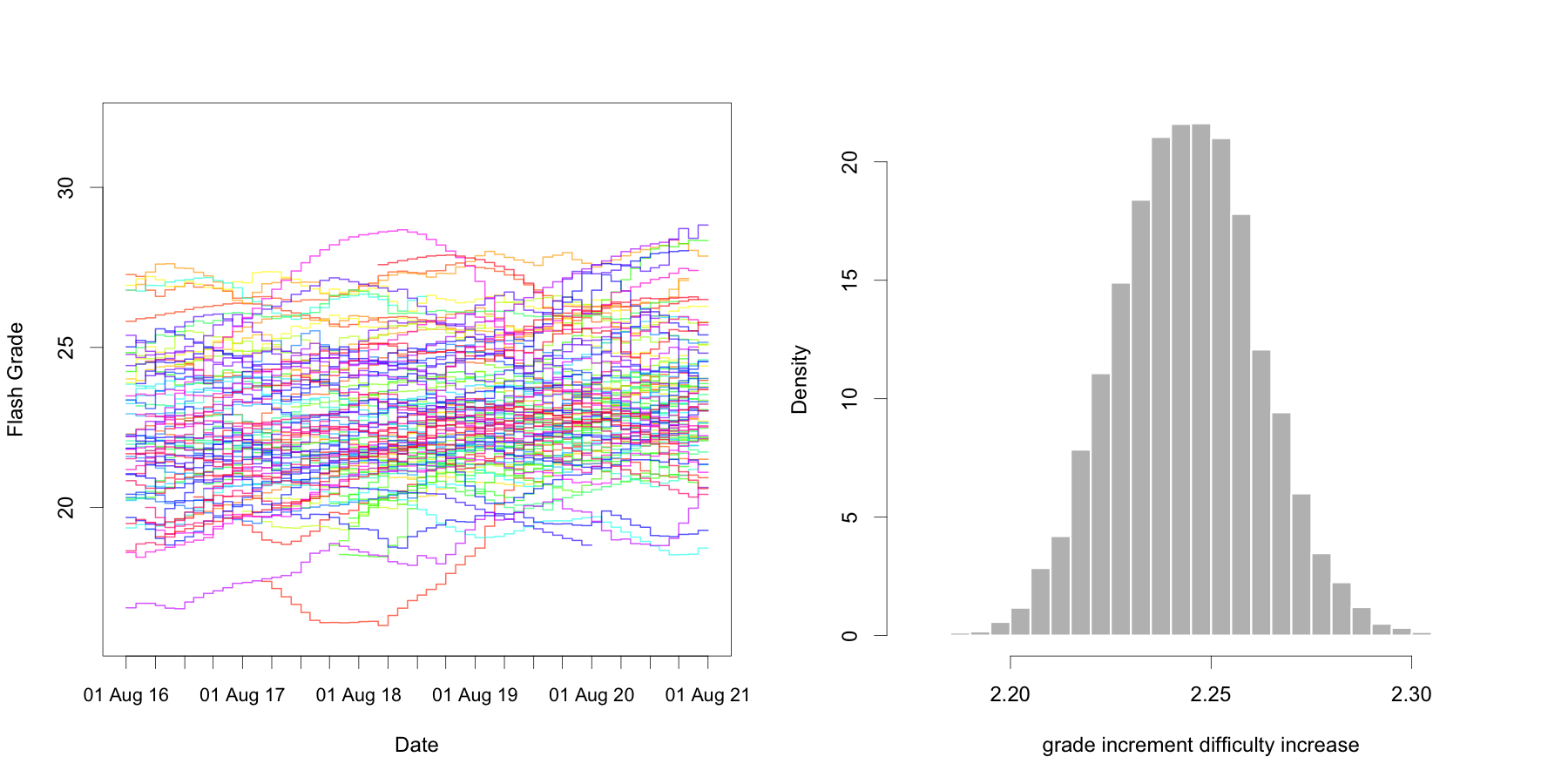}
\caption{\small The posterior estimate of each Australian climber's grade ($n=100$) through time and the posterior distribution of the proportional increase in difficulty per grade increment $d = e^m$.}
\label{aus_ascents_by_attempt}
\end{figure}

Figure \ref{aus_ascents_by_attempt} shows the estimated ``flash grade'' through time plot between August 2016 and July 2021 inclusive for 100 climbers that fulfilled our selection criteria and who climbed predominantly in Australia. The second panel reports the posterior distribution of the $e^m$, which was jointly estimated. The posterior estimate was 2.24 (95\% HPD:  $[2.21, 2.28]$).

\begin{figure}
\centering
\includegraphics[width=\textwidth]{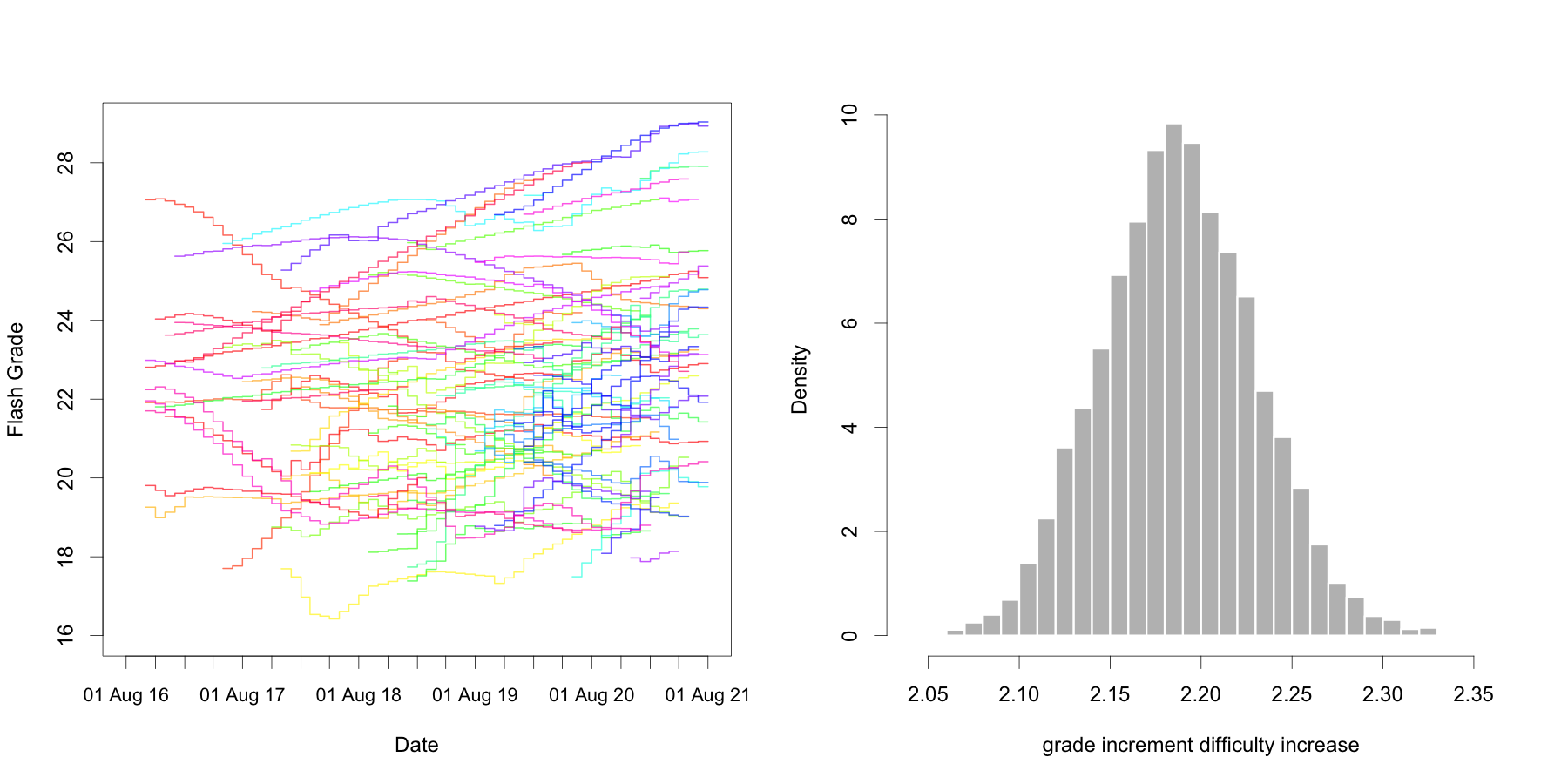}
\caption{\small The posterior estimate of each New Zealand climber's grade through time and the posterior distribution of the proportional increase in difficulty per grade increment $d = e^m$.}
\label{nz_ascents_by_attempt}
\end{figure}

Figure \ref{nz_ascents_by_attempt} shows the estimated ``flash grade'' through time plot between August 2016 and July 2021 inclusive for 89 climbers that fulfilled our selection criteria and who climbed predominantly in New Zealand. The second panel reports the posterior distribution of the $m$ parameter which was jointly estimated.

\section{Interpreting whole-community ascent success data}

In 2002, Tim O'Neill published a graph of the total number of successful ascents for each grade above V9 in Australia at the time \cite{oneill2002}. He proposed a power-law relationship between the total number of ascents achieved across the climbing community $N$ and the grade $V$. He argued that if only half the climbers, on average, can climb boulder problems when the grade is increased then that suggests such boulder problems are twice as hard. We propose to fit an exponential relationship instead:

\begin{equation}
N = e^{-rx} 
\end{equation}

One of the problems with comparing the resulting slope parameter with those estimated from whole-history data is that the number of completed boulder problems of grade $x$ is the product of the number of attempts of grade $x$ and the probability of being successful on each attempt. So the ratio of the number of successes at grade $x+1$ and the number of successes at grade $x$ is:

\begin{equation}
\frac{N_A(x+1)\Pr(\textrm{success climbing } x+1 )}{N_A(x)\Pr(\textrm{success climbing } x)}
\end{equation}

where $N_A(x)$ is the number of attempts made by the community on boulder problems of grade $x$. 
Therefore the slope of the log of total successful ascents against grade will only be the same as the $m$ parameter {\it if there has been an equal number of total attempts made by the community at each grade}. This is very unlikely to be the case at the very highest grades globally. In general we might expect that very hard problems are tried less often by the community as a whole, leading to an expectation that $r$ estimated this way would be an overestimate of $m$.

\bibliography{climbing}